\documentclass{article}

\usepackage{arxiv}

\usepackage[utf8]{inputenc} 
\usepackage[T1]{fontenc}    
\usepackage{url}            
\usepackage{booktabs}       
\usepackage{amsfonts}       
\usepackage{nicefrac}       
\usepackage{microtype}      
\usepackage{lipsum}
\usepackage{subfigure}
\usepackage{amsmath}
\usepackage{mathtools}
\usepackage[T1]{fontenc}
\usepackage{hyperref}
\usepackage{subfigure}

\title{Thin Lenses and Thin Cameras}

\author{
Wubin Pang\\
Department of Electrical and Computer Engineering\\
Duke University\\
Durham, NC 27705 \\
\texttt{pwubin28@yahoo.com}\\
\And
David J. Brady\\
Wyant College of Optical Sciences\\ University of Arizona\\
Tucson, Az 85721 \\
  \texttt{djbrady@arizona.edu}\\
}

\begin{document}

\maketitle
\begin{abstract}
 Cassegrain designs can be used to build thin lenses. We analyze the relationships between system thickness and aperture sizes of the two mirrors as well as FoV size. Our analysis shows that decrease in lens thickness imposes tight constraint on the aperture and FoV size. To mitigate this limitation, we propose to fill the gaps between the primary and the secondary with high index material. The Gassegrain optics cuts the track length into half and high index material reduces ray angle and height, consequently the incident ray angle can be increased, i.e., the FoV angle is extended. Defining telephoto ratio as the ratio of lens thickness to focal length, we achieve telephoto ratios as small as 0.43 for a visible Cassegrain thin lens and 1.20 for an infrared Cassegrain thin lens. To achieve an arbitrary FoV coverage, we present an strategy by integrating multiple thin lenses on one plane with each unit covering a different FoV region. To avoid physically tilting each unit, we propose beam steering with metasurface. By image stitching, we obtain wide FoV images. 
\end{abstract}

The need for thin cameras has progressively grown as camera applications in mobile and wearable devices has exploded over the past 20 years. Several technological advances have driven the miniaturization of camera modules, such as shrinkage of pixel pitch, complex surface figures, new plastic materials and many others. Among all factors, down-scaled pixel pitch is the main drive which also leads to downsized image sensor format and reduced sensor cost. At present, pixel size of image sensors in mobile devices is approaching $1\mu m$. Due to small pixel pitch, high angular resolution can be realized with small focal length, thus the thickness of the camera modules is reduced accordingly. Downscaled pixel size has its own tradeoffs. Firstly, small pixel size yields degraded dynamic range and increased noise level, which results in low signal to noise ratio (SNR) and leads to degenerated image quality. Moreover, the pixel size cannot be below the diffraction limited spot size. 

Recently, several groups have demonstrated thin optical elements based on metamaterials \cite{khorasaninejad2016metalenses, wang2018broadband, genevet2017recent}, geometric phase modulation \cite{escuti2016controlling} or diffractive elements \cite{wang2016chromatic}. However, while thin lenses are necessary for thin cameras, thin optics are not sufficient to produce thin optical systems. Naively, a thin lens with focal length $F$ still requires that the camera thickness be greater than $F$, no matter how thin the lens is. Previous thin camera design strategies focus on folded optics \cite{Tremblay:07, marks2013wide, tallon2016pocket}, multiaperture sampling \cite{tanida2001thin, venkataraman2013picam, shankar2008thin} and coded apertures \cite{asif2016flatcam}.

All these efforts have demonstrated useful concepts and successful ideas in producing thin imagers. Nonetheless, each one by itself also suffers from major flaws and is inadequate to create competitive thin cameras. For example, while an n-folded optic reduces track length by a factor of n, the field of view (FoV) of folded optics is very limited. This is due 
 to delicate ray path steered by multiple mirrors, where even a slight change of ray angle could result in stopped or strayed ray tracing. The field angle of a folded optics is generally less than $10^\circ$. Therefore, it is common to see folded optics being used only as telescopes for astrophotography and aerial surveillance. Other approaches include multiapture sampling strategies using super-resolution or pixel braiding for achieving desired resolution with microlens arrays of tiny focal length. Super-resolution  relies on image parallax existing on different channels. The pixel braiding strategy draws inspiration from compound eyes of insects, a pitch difference between the lenslets and the corresponding pitch is necessary and high resolution image is reconstructed through electronic stitching or pixel braiding. Despite somewhat enhanced spatial resolution, the total bandwidth product under multiaperture scheme is unsatisfactory and total pixel counts is currently less than one megapixel.

Instantaneous field of view (e.g. resolution) is fundamentally tied to focal length, specifically ${\rm ifov}=\delta /F$, where $\delta$ is the pixel pitch. This means that the goal of the system is to maximize focal length. The relative success of a thin lens method can be evaluated by using the telephoto ratio. The telephoto ratio originally is based on the telephoto design defined as the ratio between the physical length and the focal length, i.e., $L/f$. Judging from this criterion, today's miniaturized lenses barely gain any advancement in terms of reducing the telephoto ratio. Through our review on existing thin camera methods, we find the folded optics is the only approach which keeps the long focal length while doing thickness reduction. The problem of narrow FoV could be solved by multi-camera strategy, as this strategy has already become a standard measure in mobile phone industry. This letter explores optical designs that combine folded optics with camera arrays to produce thin cameras with wide FoV coverage.  

 Folded optics are traditionally used in building astronomical telescopes for achieving long focal length within limited track length. The limited FoV and severe central obscuration make it unsuitable in customer camera production where at least FoVs above $10^\circ$ are required. As cameras with compact form factor are in high demands, in 2007 Ford {\it et al.} proposed to build ultrathin cameras using folding optics and named this strategy as "origami optics"\cite{tremblay2007ultrathin,tremblay2009ultrathin}. In origami optics, the light path is reflected off annular mirrors by multiple times and the whole system are fabricated on a monolithic substrate. In comparison with the sequential layout, the thickness of the folded lenses is reduced by a factor of $n$, where $n$ denotes the number of folds.

As mentioned before, folded optics tends to have narrow FoV and small effective aperture due to obscuration. Here we discuss about Cassegrain systems. Cassegrain system is the simplest and earliest folded optics dating back to 1672. It consists of two mirrors, a primary and a secondary. The primary is a concave mirror which converges and reflects the incoming beams off to the secondary. The secondary is usually a convex mirror which reflects off the beams onto the image sensor plane. The combination of a positive primary and a negative secondary also makes a telephoto style benefiting a reduced track length even further. The occlusion of the secondary mirror results in an annular aperture which causes MTF curves to fall off over middle frequency band.  

\begin{figure}[htbp]
\centering
\includegraphics[width=5cm]{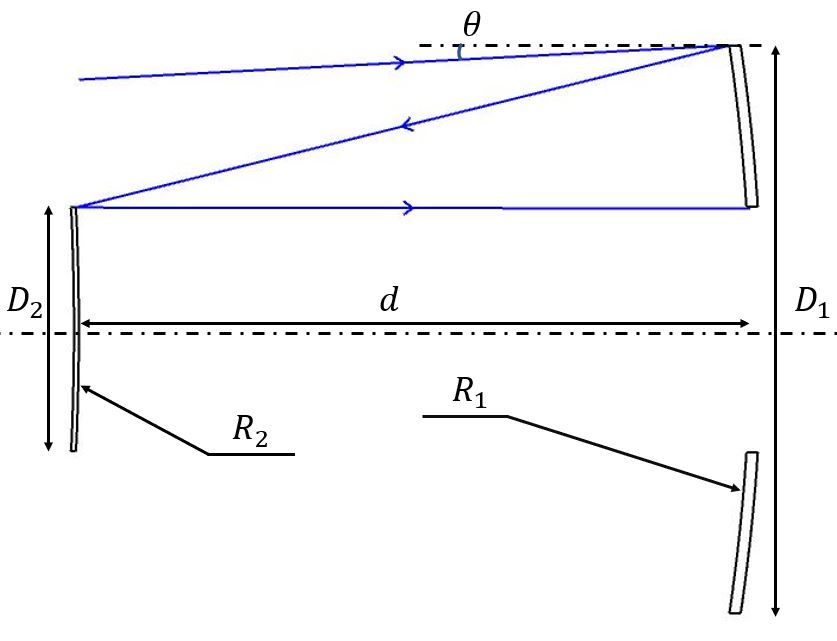}
\caption{Cross-section of a folding Cassegrain system.}
\label{fig1}
\end{figure}

Assuming a Cassegrain lens with major parameters as shown in Fig.\ref{fig1}, the radii of the curvature of the primary and the secondary are $R_1$ and $R_2$ and the aperture diameters are $D_1$ and $D_2$, respectively. The separation between the two mirrors is $d$. The top ray from the edge field angle $\theta$ is reflected off the primary, in order for the top ray to connect with the secondary, the diameter of the secondary must comply to the constraint

\begin{equation}
D_2 \geq D_1-2(\frac{D_1}{R_1}-\theta)d
\label{Eqn1}
\end{equation}

Since the diameter of the secondary cannot be larger than that of the primary, thereby the field angle $\theta < D_1/R_1$. Commonly the aperture of a mirror is far smaller than its radius of the curvature. Consequently, the upper limit for the FoV is rather small. Notably, this limitation here only accounts for the clearance of light path on the secondary, when aberration is considered as will be in real design works, the upper limit will be even lower. 

Also noticing that $D_1/R_1-\theta$ should be positive, simple rearrangement of the terms in inequation.\ref{Eqn1} relates this limitation to a constraint on the lens thickness $d$

\begin{equation}
d\geq \frac{D_1-D_2}{2(\frac{D_1}{R_1}-\theta)}
\label{Eqn2}
\end{equation}

Inequation\ref{Eqn2} explicitly shows the relationship among the lens thickness $d$, aperture diameters of the two mirrors $D_1, D_2$ and FoV angle $\theta$. Specifically, to make thin lenses or to reduce $d$, we must either increase the size of the secondary mirror $D_2$ resulting in reduced light throughput or decrease the field angle $\theta$, i.e. making narrow FoV. To summarize, the folding optics provides reduced track length $d\approx f/N$ with decreased light throughput and FoV angle.

\begin{figure}[htbp]
\centering
\includegraphics[width=4cm]{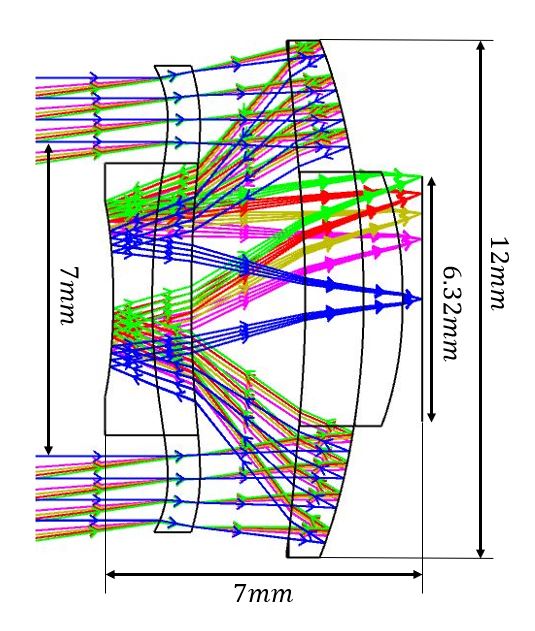}
\caption{Layout of a catadioptric miniature lens design.}
\label{fig0}
\end{figure}

\begin{figure}[htbp]
\centering
\includegraphics[width=6cm]{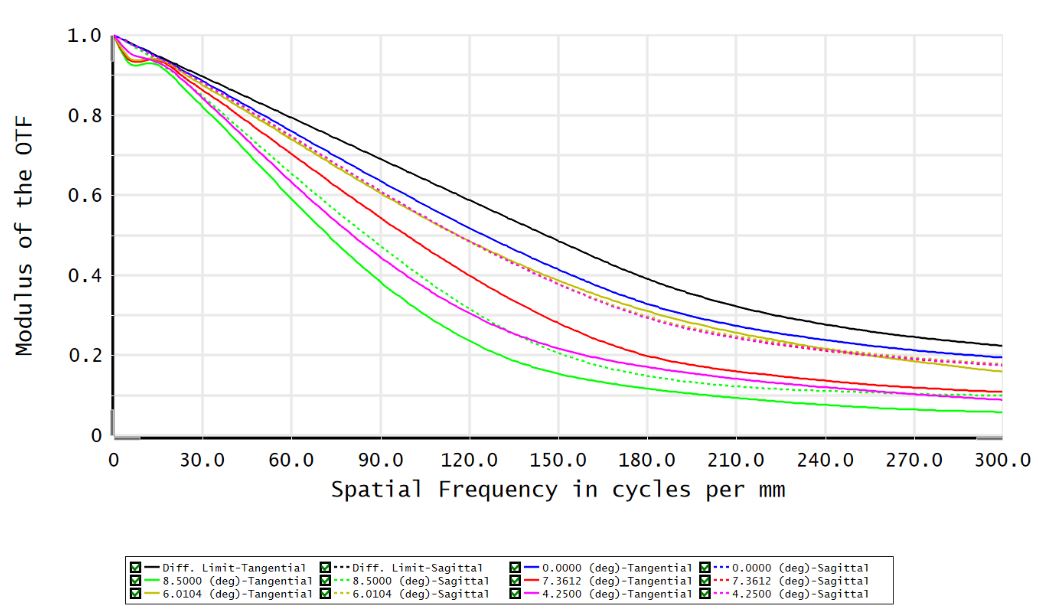}
\caption{MTF curves of the catadioptric design.}
\label{fig-1}
\end{figure}

To overcome this limitation, we can insert more optical elements and employ complex surface figures, such as the even-aspheric surface. By adding refractive elements, we create a catadioptric design style. The extra elements provide more degrees of freedom for ray path manipulation and aberration control. Here we show a design example of a miniaturized catadioptric lens aiming for miniature camera applications. For this design, we have a $20mm$ focal length, F/2 aperture, and $17^\circ$ full FoV. The working wavelength in over $486nm-656nm$ visible band. The optical layout and important dimensions are illustrated in Fig.\ref{fig0}. The design consists of two cemented doublets, each doublet has a mirror surface for ray path folding. The total track length of this design is $7mm$, which leads to a telephoto ratio of 0.35. The MTF curves are shown in Fig.\ref{fig-1}.    

Making a catadioptric design is fairly effective in increasing lens aperture and FoV. However, the achieved lens tack length (z-height) is not ideal, since a cellphone case can only accept lens z-height less than $5mm$. we still need to find approaches to reduce the telephoto ratio even further.  

\begin{figure}[htbp]
\centering
\includegraphics[width=8cm]{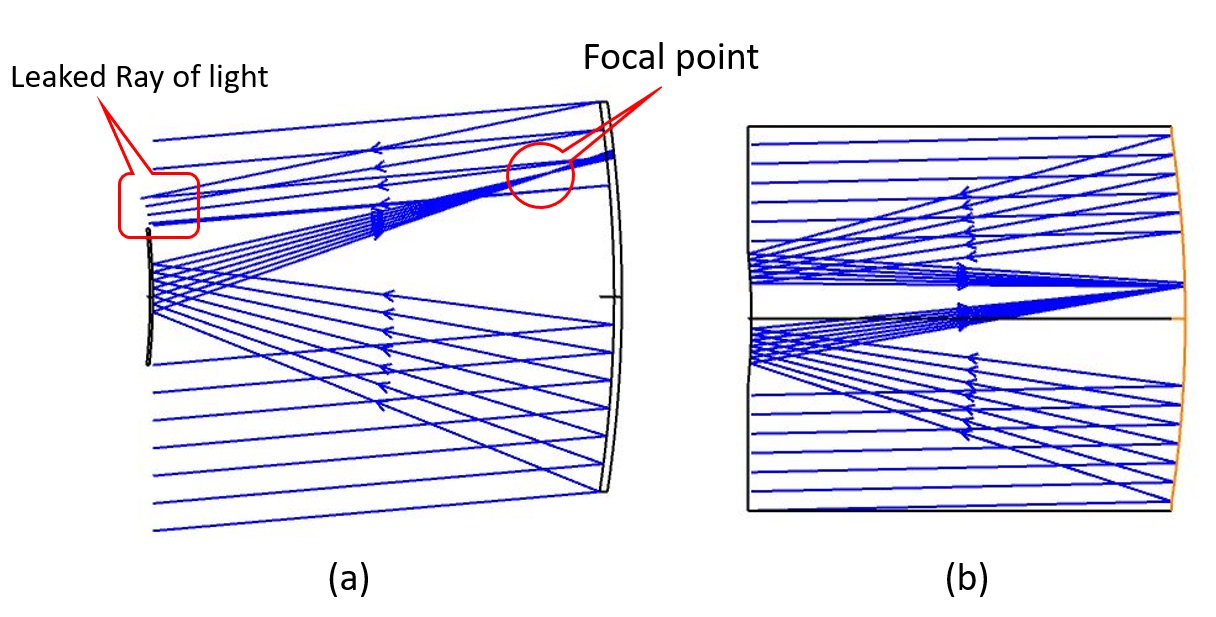}
\caption{Effective aperture size and field of view of folded lenses.(a)Without high index filling in, high angle rays are difficult to be confined or not to intersect with other rays.(b)With high index filling in, ray paths are easy to be bounded and controlled.}
\label{fig2}
\end{figure}

To solve the problem, we propose to fill the air space between the two mirrors with material of high refractive index. According to Snell's law, ray angle is inversely proportional to the refractive index of the medium $\theta \propto \frac{1}{n}$. By filling the air space with high index material, the ray angles could be reduced by a factor of $\frac{1}{n}$. As demonstrated in Fig.\ref{fig2}(a), with air gap, the growing field angle causes a portion of the ray bundle to stray away from the secondary mirror while the remaining rays interfere with the aperture of primary mirror. In contrast, as shown in Fig.\ref{fig2}(b), with high index material, the ray angle is reduced thereby ray beams trace through the system smoothly and interference free. With reduced ray angle, not only the off-axis aberration is decreased, but also the lower boundary for the thickness $d$ gets pushed to a even lower value as we replace the field angle $\theta$ with $\theta/n$ in Eqn.\ref{Eqn2}. One would worry about problems of increased weight and expense associated with the use of high index materials. But noticing that the volume of the material is proportionate to the thickness of the optics $d$, which is compressed by application of the material. Moreover, the method proposed here is targeting for miniaturized lens systems and the problems of material weight and cost are rendered insignificant.      

\begin{figure}[htbp]
    \centering
    \includegraphics[width=\linewidth]{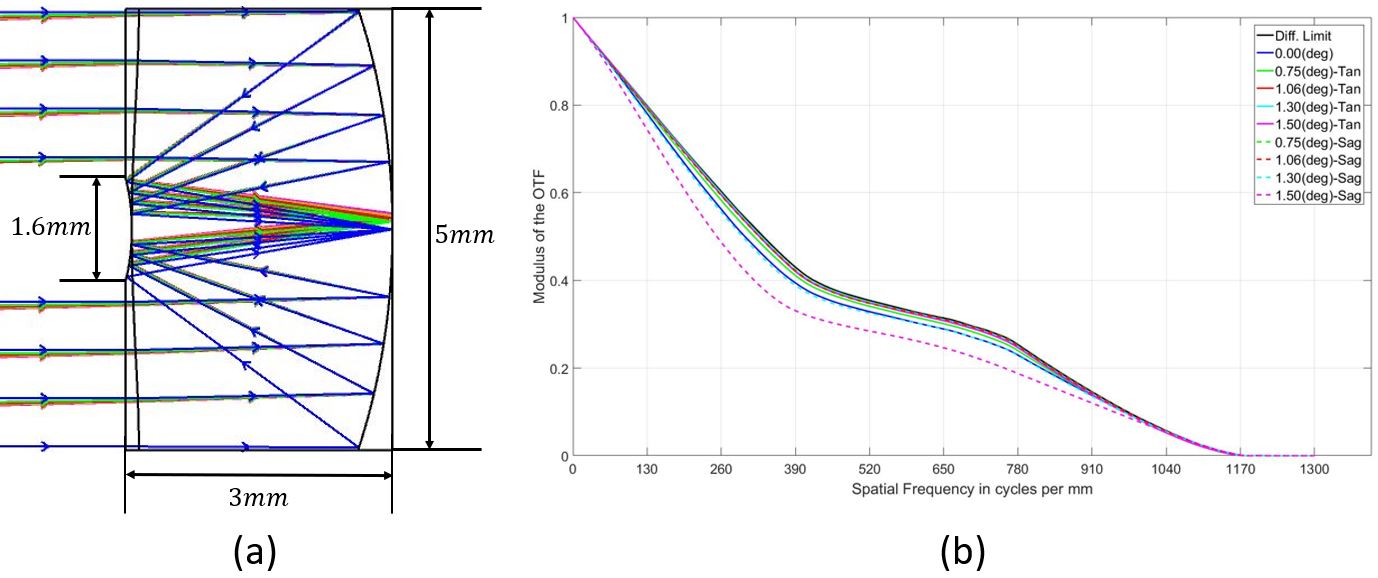}
    \caption{A Cassegrain lens design using high index material filling in.(a) Layout of the lens design.(b) MTF curves of the lens design.}
    \label{fig3}
\end{figure}

To drive home our point, we show a concrete design example in which we fill LASF35 (Nd=2.02,Vd=29.06), an  optical glass from SCHOTT catalog, into the gap between the primary and the secondary of a Cassegrain system. Fig.\ref{fig3}(a) shows the layout of the system. The two-mirror system contains three aspheric optical surfaces, the first one is refractive in the front, the others are two mirror surfaces. The image plane is in contact with the back side of the primary mirror for avoiding total internal reflection. Here are the three key specifications of the system: focal length $f=7mm$, outer aperture is $5mm$ and inner aperture is $1.6mm$, field of view: $FoV=3^\circ$. The track length of the system is $L=3mm$ resulting in a telephoto ratio of $T\approx0.43$. For the annular aperture, the outer diameter is $5mm$, the inner diameter is $1.6mm$, hence the effective diameter is $4.74mm$. The modulation transfer function (MTF) curves measuring the imaging performance are shown in Fig.\ref{fig3}(b). The system achieves diffraction limited performance. Because of the fill into of high index material,the cut-off frequency lies at about 1170cycles/mm. 
 
\begin{figure}[htbp]
\centering
\includegraphics[width=\linewidth]{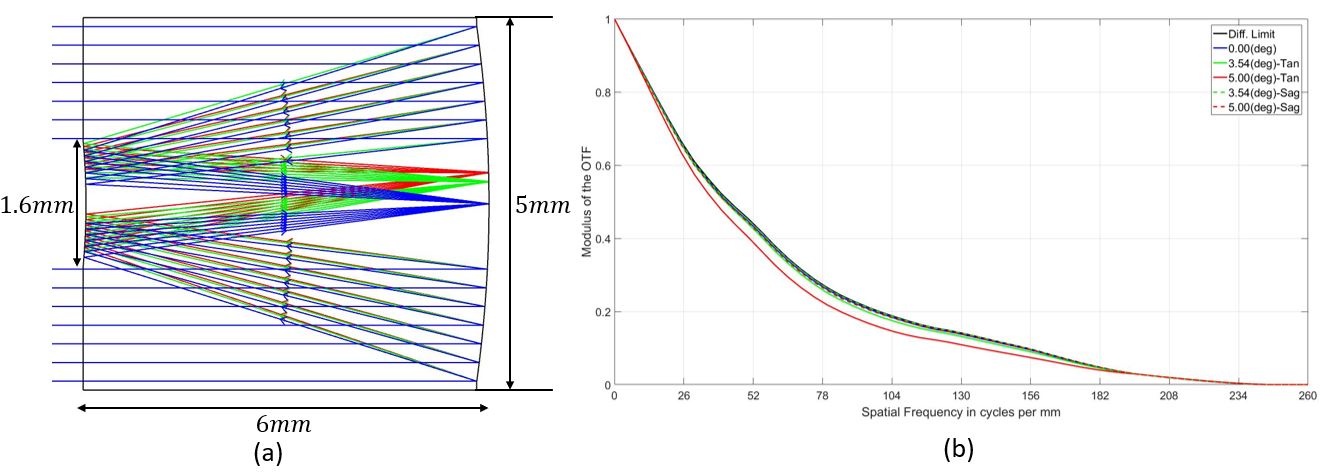}
\caption{A infrared Cassegrain system with a substrate made of germanium.(a)Layout of the lens design.(b) MTF curves of the lens design.}
\label{fig4}
\end{figure}

The index of refraction of natural materials can rarely reach 2 for visible spectral region, while it can be above 4 for the infrared region. Naturally, we apply our method in an lens design for infrared imagers. In our second example, a Cassegrain system is used in infrared imaging. The gap between the primary and the secondary is filled with germanium $(n=4.0243\ {\rm at}\ \lambda=4\mu m\ and\ n=4.0032\ {\rm at}\ \lambda=10\mu m)$. The layout of the system is shown in Fig.\ref{fig4}(a). The focal length is $5mm$, outer aperture is $5mm$, the obscuration diameter is $1.6mm$ and the field of view (FoV) is 10 degrees, the total track length is $6mm$ which gives a telephoto ratio of $1.27$. The MTF curves in Fig.\ref{fig4}(b) show diffraction limited performance.             

For normal optical materials, high refractive index often indicates high dispersion and absorption. New metamaterials can generate artificial high index over a wide spectrum band with well controlled dispersion and reasonable light efficiency\cite{shrestha2018broadband}. High reflectivity on the interface of high index difference could be solved by anti-reflection coatings.     
 
Undoubtedly, high index material filling also adds to the weight and material cost for the imagers. This is impractical for large folded lenses and unacceptable for huge astronomical telescopes which accounts for most usage of folded lenses currently. However, as we bring folded lenses into developing miniaturized cameras for mobile platforms, the side effects of extra weight and cost are totally insignificant.

In smartphones,  major manufactures increasingly compete to integrate more camera units in their devices. Array cameras are also popular in security, surveillance, automation, virtual reality (VR) and augmented reality (AR). Lens arrays achieve wide angle FoV by pointing each lens unit to a different viewing angle and produce panoramas by stitching sub-image from different channels. Currently steering lens units requires physical tilt so that the optical axis is aiming at the targeted viewing direction. For a thin camera of greater lateral dimension than longitudinal dimension, tilting the camera could cause increase in thickness and difficulty in system assembling. As shown in the US. patent\cite{marks2016apparatus}, a part of the aperture is clipped off for preserving the required array thickness. Alternatively, a curved mount could be implemented as shown in many other practices. Lens tilting complicates the mechanical design as well as assembly procedure, sometimes also adds extra thickness to the array.

To avoid all the inconvenience associated with lens tilting, method of granting the camera units with deflected view is required. One would imagine that glass wedges could do the work by changing direction of the incoming beam of light. However, glass wedges incur chromatic aberration and also add on significant extra thickness. Diffracting elements can also be used to skew the view, but also introduce chromatic aberration as well as decrease light efficiency as light is being divided into different orders. Meanwhile, metasurface, as a hot research topic in which considerable progress has been made, provides a solution for pointing camera units in different directions without having to physically tilt the camera unit.

The advancement in matesurfaces and high index material of wide band and low dispersion offers new opportunities for the thin lens research and development. In order to tilt the optical axis of a lens, a linear phase modulation should be applied to the incoming light. A meta-surface with phase modulation function expressed as $e^{i\frac{2\pi}{\lambda} (\alpha x+\beta y)}$ would bend a skewed ray with direction vector $(\alpha,\beta,\sqrt{1-\alpha^2-\beta^2})$ into the new direction in parallel with the optical axis with direction vector as $(0,0,1)$. Since the modulation is linear, it does not introduce any aberrations.

\begin{figure}[htbp]
 \centering
 \includegraphics[width=8cm]{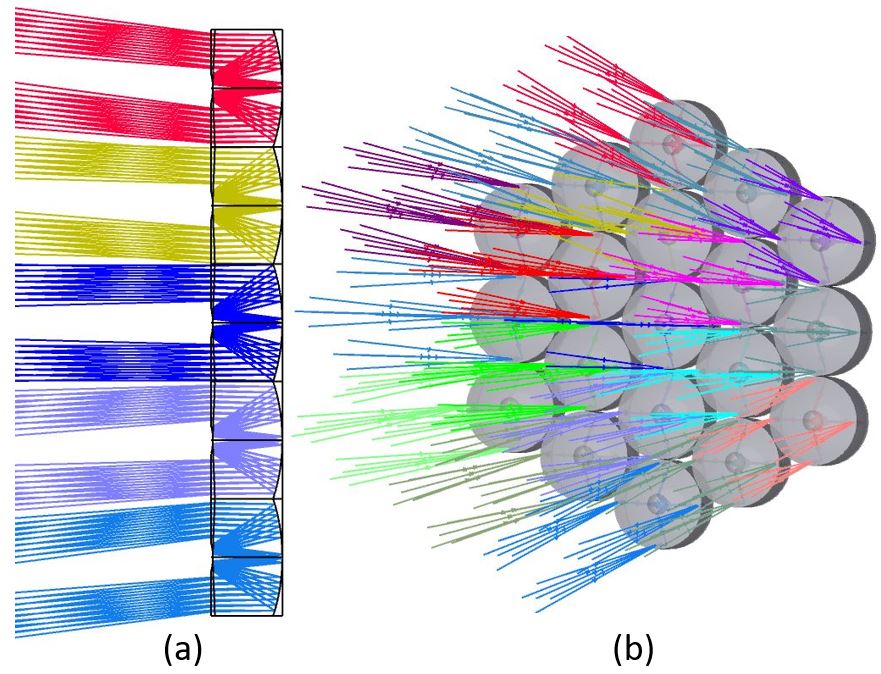}
  \caption{Meta-pointers.(a) Metasurface make a column of lens arrays to point to different directions without having to tilt the constituent lens units. (b) A 2D lens array of flat formation achieves extensive field of view using meta-surface.}
  \label{fig5}
\end{figure}

"Meta tilting" is especially advantageous for systems of being wider in lateral dimension than in axial dimension. This is the case for our Cassegrain Visible system shown in fig.\ref{fig3}. As shown in Fig.\ref{fig5}(a), an row of five lens units achieves a total $15^\circ\times 3^\circ$ field of view with a specified phase modulation simulating a meta-surface attached to the first surface of each lens unit. The Fig.\ref{fig5}.(b) illustrates an extension from one row to a two dimensional array of cameras consisting of nineteen camera units in a triangular formation. These metasurfaces can be made as detachable patches, therefore, the center viewing direction of each channel can be changed by replacing the front metasurface patch easily.

To summarize, thin lenses are desirable in many application scenarios. They extends the accessibility of cameras, reduces camera size and weight and facilitates portable and wearable devices. Currently, majority of thin lenses are only a result from down-scaled normal sized lenses, which is mainly enabled by smaller semiconductor pixel pitch and sensor formats. Though this leads to a shortened track length, it also results in decreased transverse dimensions and reduced light flux.  Negative effects include reduced dynamic range and degraded SNR, with which image quality suffers. Therefore, superior solutions for thin lens should seek to preserve the lens focal length and aperture size. Folded optics represented by Cassegrain system provides a valid solution for building compact camera modules with long focal length and large aperture size. The problems of narrow FoV and aperture obcuration associated with the folded optics can be mitigated by the employment of high index materials. By filling the air gap with high index material, beams of light can be easily confined inside the system space and ray blockage and stray light is effective prevented. The two Cassegrain systems in our design example section, one in visible region and the other in infrared region, demonstrate the effectiveness of our approach. Ultimately, to be able to realize thin cameras of real wide FoV, we suggest employing a simplest possible metasurface which only needs to generate linear phase modulations. By attaching or clipping such a surface to the front of a imaging system, the boresight of the optics can be changed. When our suggested metasurface is combined with array cameras, all the array units could be assembled on a flat formation with parallel axial orientations. This tilt free camera steering method essentially provides a way to building wide angle array cameras without having to surrender the accomplished thickness in each constituent unit.  
\bibliographystyle{unsrt}  
\bibliography{sample}

\end{document}